\documentclass[12pt]{article}

\usepackage{graphicx}
\usepackage{cite}
\usepackage{color}

\newcommand{\nn}{\nonumber}

\def\beq{\begin{equation}}
\def\eeq{\end{equation}}
\def\nn{\nonumber}
\def\bea{\begin{eqnarray}}
\def\eea{\end{eqnarray}}
\def\ba{\begin{array}}                  
\def\ea{\end{array}}
\newcommand{\FUTA}{{\bf FUTA}}
\newcommand{\FUTB}{{\bf FUTB}}

\def\reffi#1{\mbox{Fig.~\ref{#1}}}

\definecolor{Orange}{named}{Orange}
\definecolor{Purple}{named}{Purple}

\graphicspath{{figs/}}

\begin{document}

\thispagestyle{empty}
\setcounter{page}{0}
\def\thefootnote{\fnsymbol{footnote}}
\vspace*{-3.0cm}

\begin{flushright}
CERN-PH-TH/2012-011
\end{flushright}

\vspace{1cm}

\begin{center}

{\fontsize{15}{1} 
\sc {\bf Finite Unified Theories and the Higgs-Boson}}

\vspace{1cm}

{\sc 
Sven Heinemeyer$^1$
\footnote{
email: Sven.Heinemeyer@cern.ch
}, %
 Myriam Mondrag\'on$^2$
\footnote{
email: myriam@fisica.unam.mx
} \\%
~and George Zoupanos$^3$
\footnote{
email: George.Zoupanos@cern.ch
}\footnote{On leave from Physics Department,
 National Technical University,
Zografou Campus: Heroon
    Polytechniou 9, 15780 Zografou, Athens, Greece}
}

\vspace*{0.5cm}
 
$^1$\sl Instituto de F\'{\i}sica de Cantabria (CSIC-UC)\\
\sl Edificio Juan Jorda,
Avda. de Los Castros s/n \\
\sl 39005 Santander, Spain
\\[3mm]
$^2$ 
\sl Instituto de F\'{\i}sica\\
\sl Universidad Nacional Aut\'onoma de M\'exico\\
\sl Apdo. Postal 20-364, M\'exico 01000 D.F., M\'exico
\\[3mm]
$^3$ \sl Theory Group, Physics Department\\
\sl CERN, Geneva, Switzerland
 
\end{center}

\vspace{0.3cm}

\begin{abstract}
All-loop Finite Unified Theories (FUTs) are very interesting $N = 1$
supersymmetric Grand Unified Theories (GUTs) realising an old field
theory dream, and moreover have a remarkable predictive power due to
the required reduction of couplings. 
Based on
this theoretical framework phenomenologically consistent FUTs
have been constructed. Here we review two FUT models based on the
$SU(5)$ gauge group, which can be seen as special, restricted and thus very
predictive versions of the MSSM. 
We show that from the requirement of correct prediction of quark masses and
other experimental constraints a light Higgs-boson mass in the range
$M_h \sim 121 - 126$~GeV is predicted, in striking agreement with recent
experimental results from ATLAS and CMS. The model furthermore naturally
predicts a relatively heavy spectrum with colored supersymmetric particles
above $\sim 1.5$~TeV in agreement with the non-observation of those particles
at the LHC. 

\end{abstract}

\newpage

\section{Introduction}

The success of the Standard Model (SM) of Elementary Particle Physics
is seriously limited by the presence of a plethora of free
parameters. An even more disturbing fact is that the best bet for
Physics beyond the SM namely the minimal supersymmetric extension of
the SM (MSSM), which is expected to bring us one step further towards
a more fundamental understanding of Nature, introduces around a
hundred additional free parameters. 
To reduce the number of free parameters of a theory, and thus render
it more predictive, one is usually led to introduce a symmetry.  Grand
Unified Theories (GUTs) are very good examples of such a procedure
\cite
{Pati:1973rp,Georgi:1974sy,Georgi:1974yf,Fritzsch:1974nn,Carlson:1975gu}.
For instance, in the case of minimal $SU(5)$, because of (approximate)
gauge coupling unification, it was possible to reduce the gauge
couplings by one and give a prediction for one of them. 
In fact, LEP data \cite{Amaldi:1991cn} 
seem to
suggest that a further symmetry, namely $N=1$ global supersymmetry (SUSY) 
\cite{Dimopoulos:1981zb,Sakai:1981gr} 
should also be required to make the prediction viable.
GUTs can also
relate the Yukawa couplings among themselves, again $SU(5)$ provided
an example of this by predicting the ratio $M_{\tau}/M_b$
\cite{Buras:1977yy} in the SM.  Unfortunately, requiring
more gauge symmetry does not seem to help, since additional
complications are introduced due to new degrees of freedom, in the
ways and channels of breaking the symmetry, and so on.

A natural extension of the GUT idea is to find a way to relate the
gauge and Yukawa sectors of a theory, that is to achieve Gauge-Yukawa
Unification (GYU) \cite{Kubo:1995cg,Kubo:1997fi,Kobayashi:1999pn}.  A
symmetry which naturally relates the two sectors is supersymmetry, in
particular $N=2$ SUSY \cite{Fayet:1978ig}.  It turns out, however, that $N=2$
supersymmetric theories have serious phenomenological problems due to
light mirror fermions.  Also in superstring theories and in composite
models there exist relations among the gauge and Yukawa couplings, but
both kind of theories have phenomenological problems, which we are not
going to address here.

Finite Unified Theories (FUTs) are $N=1$ supersymmetric Grand Unified
Theories (GUTs) which can be made finite to all-loop orders, including
the soft supersymmetry breaking sector.  The constructed {\it finite
  unified} $N=1$ supersymmetric SU(5) GUTs predicted correctly from
the dimensionless sector (Gauge-Yukawa unification), among others, the
top quark mass \cite{Kapetanakis:1992vx,Mondragon:1993tw}.
Eventually, the full theories can be made all-loop finite and their
predictive power is extended to the Higgs sector and the s-spectrum
\cite{Heinemeyer:2007tz}. For a detailed discussion see
\cite{Kubo:1997fi,Kobayashi:2001me,Heinemeyer:2010xt}. Here we limit
ourselves to a brief review.

Consider  a chiral, anomaly free,
$N=1$ globally supersymmetric
gauge theory based on a group G with gauge coupling
constant $g$. The
superpotential of the theory is given by
\begin{equation}
 W= \frac{1}{2}\,m^{ij} \,\Phi_{i}\,\Phi_{j}+
\frac{1}{6}\,C^{ijk} \,\Phi_{i}\,\Phi_{j}\,\Phi_{k}~, 
\label{1}
\end{equation}
where $m^{ij}$ (the mass terms) and $C^{ijk}$ (the Yukawa couplings)
are gauge invariant tensors, and the matter field $\Phi_{i}$
transforms according to the irreducible representation $R_{i}$ of the
gauge group $G$.  All the one-loop $\beta$-functions of the theory
vanish if the $\beta$-function of the gauge coupling $\beta_g^{(1)}$,
and the anomalous dimensions of the Yukawa couplings
$\gamma_i^{j(1)}$, vanish, i.e.
\begin{equation}
\sum _i \ell (R_i) = 3 C_2(G) \,,~
\frac{1}{2}C_{ipq} C^{jpq} = 2\delta _i^j g^2  C_2(R_i)\ ,
\label{2}
\end{equation}
where $\ell (R_i)$ is the Dynkin index of $R_i$, and $C_2(G)$ is the
quadratic Casimir invariant of the adjoint representation of $G$. A
theorem given in
\cite{Lucchesi:1987he,Piguet:1986td,Piguet:1986pk,Lucchesi:1996ir}
then guarantees the vanishing of the $\beta$-functions to all-orders
in perturbation theory.  This requires that, in addition to the
one-loop finiteness conditions (\ref{2}), the Yukawa couplings are
reduced in favour of the gauge
coupling~\cite{Zimmermann:1984sx,Oehme:1984yy,Ma:1977hf,Ma:1984by,Chang:1974bv,Nandi:1978fw}.

In the soft breaking sector,
the one- and two-loop
finiteness for the trilinear terms $h^{ijk}$ can be achieved by \cite{Jack:1994kd}
\beq
 h^{ijk} = -M C^{ijk}+\dots =-M
\rho^{ijk}_{(0)}\,g+O(g^5)~.
\label{eq:hMC}
\eeq
It was also found that the soft supersymmetry
breaking (SSB) scalar masses in Gauge-Yukawa and finite unified models
satisfy a sum rule \cite{Kobayashi:1998jq,Kubo:1994bj}
\begin{equation}
\frac{(~m_{i}^{2}+m_{j}^{2}+m_{k}^{2}~)}{M M^{\dag}} =
1+\frac{g^2}{16 \pi^2}\,\Delta^{(2)}
+O(g^4)~
\label{sumr}
\end{equation}
for i, j, k, where $\Delta^{(2)}$ is
the two-loop correction, which vanishes when all the soft scalar
masses are the same at the unification point.

\section{\boldmath{$SU(5)$}  Finite Unified Theories}

We will examine here all-loop Finite Unified theories
with $SU(5)$ gauge group, where the reduction of couplings has been
applied to the third generation of quarks and leptons.  An extension
to three families, and the generation of quark mixing angles and
masses in Finite Unified Theories has been addressed in
\cite{Babu:2002in}, where several examples are given. These
extensions are not considered here. 
The particle content of the models we will study consists of the
following supermultiplets: three ($\overline{\bf 5} + \bf{10}$),
needed for each of the three generations of quarks and leptons, four
($\overline{\bf 5} + {\bf 5}$) and one ${\bf 24}$ considered as Higgs
supermultiplets. 
When the gauge group of the finite GUT is broken the theory is no
longer finite, and we will assume that we are left with the MSSM.

A predictive Gauge-Yukawa unified $SU(5)$ model which is finite to all
orders, should 
have the following properties:
\begin{enumerate}
\item 
One-loop anomalous dimensions are diagonal,
i.e.,  $\gamma_{i}^{(1)\,j} \propto \delta^{j}_{i} $.
\item Three fermion generations, in the irreducible representations
  $\overline{\bf 5}_{i},{\bf 10}_i~(i=1,2,3)$, which obviously should
  not couple to the adjoint ${\bf 24}$.
\item The two Higgs doublets of the MSSM should mostly be made out of a
pair of Higgs quintet and anti-quintet, which couple to the third
generation.
\end{enumerate}

In the following we discuss two versions of the all-order finite
model.  The model of ref.~\cite{Kapetanakis:1992vx,Mondragon:1993tw},
which will be labeled ${\bf A}$, and a slight variation of this model
(labeled ${\bf B}$), which can also be obtained from the class of the
models suggested in ref.~\cite{Kazakov:1995cy} with a modification to
suppress non-diagonal anomalous dimensions.

The  superpotential which describes the two models 
takes the form \cite{Kapetanakis:1992vx,Mondragon:1993tw,Kobayashi:1997qx}
\bea
W &=& \sum_{i=1}^{3}\,[~\frac{1}{2}g_{i}^{u}
\,{\bf 10}_i{\bf 10}_i H_{i}+
g_{i}^{d}\,{\bf 10}_i \overline{\bf 5}_{i}\,
\overline{H}_{i}~] \nn \\\nn
&+&g_{23}^{u}\,{\bf 10}_2{\bf 10}_3 H_{4} 
  +g_{23}^{d}\,{\bf 10}_2 \overline{\bf 5}_{3}\,
\overline{H}_{4}+
g_{32}^{d}\,{\bf 10}_3 \overline{\bf 5}_{2}\,
\overline{H}_{4} \\
&+&\sum_{a=1}^{4}g_{a}^{f}\,H_{a}\, 
{\bf 24}\,\overline{H}_{a}+
\frac{g^{\lambda}}{3}\,({\bf 24})^3~,
\label{zoup-super}
\eea
where 
$H_{a}$ and $\overline{H}_{a}~~(a=1,\dots,4)$
stand for the Higgs quintets and anti-quintets.

The main difference between model ${\bf A}$ and model ${\bf B}$ is
that two pairs of Higgs quintets and anti-quintets couple to the ${\bf
  24}$ in ${\bf B}$, so that it is not necessary to mix them with
$H_{4}$ and $\overline{H}_{4}$ in order to achieve the triplet-doublet
splitting after the symmetry breaking of $SU(5)$
\cite{Kobayashi:1997qx}.  Thus, although the particle content is the
same, the solutions to the finiteness equations and the sum rules are
different, which will reflect in the phenomenology.

The non-degenerate and isolated solutions to $\gamma^{(1)}_{i}=0$ for
 model \FUTA, which are the boundary conditions for the Yukawa
 couplings at the GUT scale, are: 
\bea 
&& (g_{1}^{u})^2
=\frac{8}{5}~g^2~, ~(g_{1}^{d})^2
=\frac{6}{5}~g^2~,~
(g_{2}^{u})^2=(g_{3}^{u})^2=\frac{8}{5}~g^2~,\label{zoup-SOL5}\\
&& (g_{2}^{d})^2 = (g_{3}^{d})^2=\frac{6}{5}~g^2~,~
(g_{23}^{u})^2 =0~,~
(g_{23}^{d})^2=(g_{32}^{d})^2=0~,
\nonumber\\
&& (g^{\lambda})^2 =\frac{15}{7}g^2~,~ (g_{2}^{f})^2
=(g_{3}^{f})^2=0~,~ (g_{1}^{f})^2=0~,~
(g_{4}^{f})^2= g^2~.\nonumber 
\eea 
In the dimensionful sector, the sum rule gives us the following
boundary conditions at the GUT scale for this model
\cite{Kobayashi:1997qx,Jones:1984qd,Leon:1985jm}: 
\bea
m^{2}_{H_u}+
2  m^{2}_{{\bf 10}} &=&
m^{2}_{H_d}+ m^{2}_{\overline{{\bf 5}}}+
m^{2}_{{\bf 10}}=M^2 ~~,
\label{sumrA}
\eea
and thus we are left with only three free parameters, namely
$m_{\overline{{\bf 5}}}\equiv m_{\overline{{\bf 5}}_3}$, 
$m_{{\bf 10}}\equiv m_{{\bf 10}_3}$
and $M$.

For the model \FUTB\ the non-degenerate and isolated solutions to
$\gamma^{(1)}_{i}=0$ give us: 
\bea 
&& (g_{1}^{u})^2
=\frac{8}{5}~ g^2~, ~(g_{1}^{d})^2
=\frac{6}{5}~g^2~,~
(g_{2}^{u})^2=(g_{3}^{u})^2=\frac{4}{5}~g^2~,\label{zoup-SOL52}\\
&& (g_{2}^{d})^2 = (g_{3}^{d})^2=\frac{3}{5}~g^2~,~
(g_{23}^{u})^2 =\frac{4}{5}~g^2~,~
(g_{23}^{d})^2=(g_{32}^{d})^2=\frac{3}{5}~g^2~,
\nonumber\\
&& (g^{\lambda})^2 =\frac{15}{7}g^2~,~ (g_{2}^{f})^2
=(g_{3}^{f})^2=\frac{1}{2}~g^2~,~ (g_{1}^{f})^2=0~,~
(g_{4}^{f})^2=0~,\nonumber 
\eea 
and from the sum rule we obtain:
\beq
m^{2}_{H_u}+
2  m^{2}_{{\bf 10}} =M^2~,~
m^{2}_{H_d}-2m^{2}_{{\bf 10}}=-\frac{M^2}{3}~,
m^{2}_{\overline{{\bf 5}}}+
3m^{2}_{{\bf 10}}=\frac{4M^2}{3}~,
\label{sumrB}
\eeq
i.e., in this case we have only two free parameters  
$m_{{\bf 10}}\equiv m_{{\bf 10}_3}$  and $M$.

\section{Predictions of  the  \boldmath{$SU(5)$} models}

We confront now  the predictions of the four models, \FUTA\ and \FUTB, each
with $\mu < 0$ and $\mu > 0$, 
with the experimental data, starting with the heavy quark masses (see
refs.~\cite{Heinemeyer:2010xt,Heinemeyer:2007tz} for more details).
Since the gauge symmetry is spontaneously broken below $M_{\rm GUT}$,
the finiteness conditions do not restrict the renormalization
properties at low energies, and all it remains are boundary conditions
on the gauge and Yukawa couplings (\ref{zoup-SOL5}), the $h=-MC$
relation eq.(\ref{eq:hMC}), and the soft scalar-mass sum rule
(\ref{sumr}) at $M_{\rm GUT}$, as applied in each of the models.  Thus we
examine the evolution of these parameters according to their RGEs up
to two-loops for dimensionless parameters and at one-loop for
dimensionful ones with the relevant boundary conditions.  Below
$M_{\rm GUT}$ their evolution is assumed to be governed by the MSSM.
We further assume a unique supersymmetry breaking scale $M_{s}$ and therefore
below that scale the effective theory is just the SM.

As a first step, we compare the predictions of the two models, \FUTA\
and \FUTB\ (both with $\mu >0$ and $\mu <0$), with the experimental values
of the top and bottom quark masses.  We use for the top quark the
value for the pole mass \cite{:2009ec}%
\footnote{Using the most up-to-date value of 
$M_t^{\rm exp} = (173.2 \pm 0.9) ~{\rm GeV}$~\cite{Lancaster:2011wr} would
  have a minor impact on our analysis.}
\beq 
M_t^{\rm exp} = (173.1 \pm 1.3) ~{\rm GeV}~,
\eeq 
where we notice that the theoretical values for $M_t$ may
suffer from a correction of $\sim 4 \%$ \cite{Kubo:1997fi}.  For the
bottom quark mass we use the value at $M_Z$ \cite{Nakamura:2010zzi} 
\beq
m_b(M_Z) = (2.83 \pm 0.10) ~{\rm GeV} 
\eeq 
to avoid uncertainties that come from the
furhter running from the $M_Z$ to the $m_b$ mass, and where we have
taken the $\Delta_b$ effects into account~\cite{deltab1}.

\begin{figure}
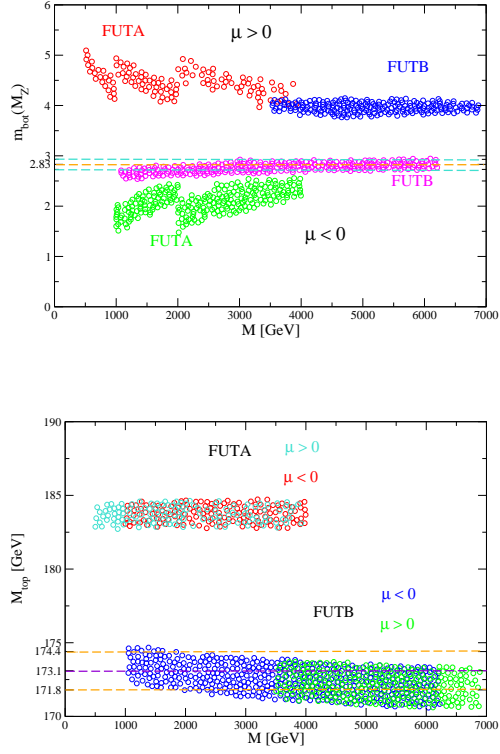

\begin{center}
{\includegraphics[width=6.5cm]{MvsMBOT-light2}\\
\vspace{1cm}
 \includegraphics[width=6.5cm]{MvsMTOP-light2}}
\caption{$m_b(M_Z)$ as function of $M$ (top) and 
$M_{\rm top}$  as function of $M$ (bottom) for models $\bf FUTA$ and
$\bf FUTB$, for $\mu<0 $ and $\mu >0$.}
\label{fig:MQvsM}
\end{center}
\end{figure}

From \reffi{fig:MQvsM} it is clear that the model \FUTB\ with $\mu <0$ is
the only one where both top and bottom quark masses lie within
experimental limits.  In this case the value of $\tan \beta$ is found to
be $\sim 48$.
Thus, we will concentrate now on the results for \FUTB, $\mu < 0$.

In the case where all the soft scalar masses are universal at the
unfication scale, there is no region of $M$ below ${\cal O}$(few~TeV)
in which $m_{\tilde \tau} > m_{\chi^0}$ is satisfied (where $m_{\tilde
  \tau}$ is the lightest $\tilde \tau$ mass, and $m_{\chi^0}$ the
lightest neutralino mass). However, this problem can be solved naturally,
thanks to the sum rule (\ref{sumrB}), see 
Refs.~\cite{Heinemeyer:2007tz,Heinemeyer:2010xt} for details.
A related problem concerns the agreement of our predictions with the cold dark
matter constraints. Again a detailed discussion can be found in 
Refs.~\cite{Heinemeyer:2007tz,Heinemeyer:2010xt}.

\medskip
As a second step, we impose the conditions of successful radiative
electroweak symmetry breaking, $m_{\tilde\tau}^2 > 0$ and
$m_{\tilde\tau}> m_{\chi^0}$. We furthermore require agreement at the
95\%~C.L. with constraints coming from $B$-physics, 
namely the experimental bounds on $\rm{BR} (b \to s \gamma)$~\cite{HFAG}
and $\rm{BR} (B_s \to \mu^+\mu^-)$~\cite{CMSLHCb} (which we have evaluated
with Micromegas \cite{Belanger:2001fz}). This way, we find a prediction for
the lightest Higgs mass and the SUSY spectra.  From the analysis we find
that the lightest observable particle (LOSP) is either the stau or the
second lightest neutralino, with mass starting around $\sim 500$ GeV.

\begin{table}[htb!]
\begin{center}
\renewcommand{\arraystretch}{1.3} 
\caption{A representative spectrum of a light {\bf FUTB}, $\mu <0$ spectrum.
\hfill\mbox{}}
{\begin{tabular}{|l|l||l|l|}
\hline 
$m_b$($M_Z$) &  2.71 GeV &
$M_t$ &    172.2 GeV\\ \hline
$M_h$ &  123.1 GeV & 
$M_A$ &  680 GeV\\ \hline 
$M_H$ &  679 GeV& 
$M_{H^\pm}$ &  685 GeV \\ \hline 
Stop1 &  1876 GeV &
Stop2 &    2146 GeV \\ \hline
Sbot1 &   1849 GeV & 
Sbot2 &    2117 GeV\\ \hline 
Mstau1 &    635 GeV & 
Mstau2 &    867 GeV\\ \hline 
Char1 &    1072 GeV & 
Char2 &    1597 GeV\\ \hline
Neu1  &    579 GeV &
Neu2  &    1072 GeV \\ \hline 
Neu3  &    1591 GeV &
Neu4  &    1596 GeV \\ \hline
 M1 &    580 GeV& 
 M2 &   1077 GeV\\ \hline 
 Gluino &    2754 GeV & 
&  \\
\hline 
\end{tabular}
\renewcommand{\arraystretch}{1.0}
\label{table:mass}
}
\end{center}
\end{table}

\begin{figure}[htb!]
\vspace{1cm}
\centerline{\includegraphics[width=8cm]{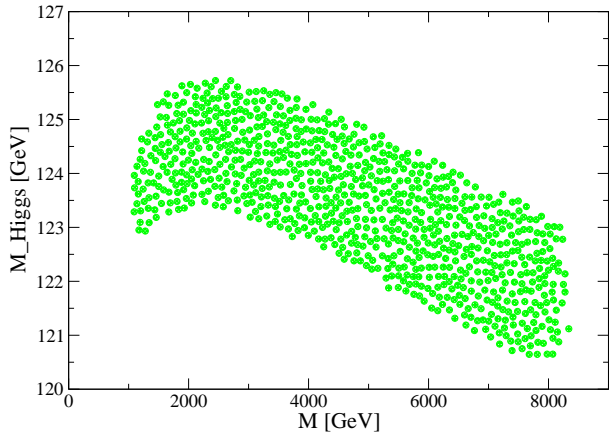}}
\caption{The lighest Higgs mass $M_h$ as function of the unified gaugino mass
  $M$.} 
\label{fig:Higgs}
\end{figure}

\begin{figure}[htb!]
           \centerline{\includegraphics[width=10cm,angle=0]{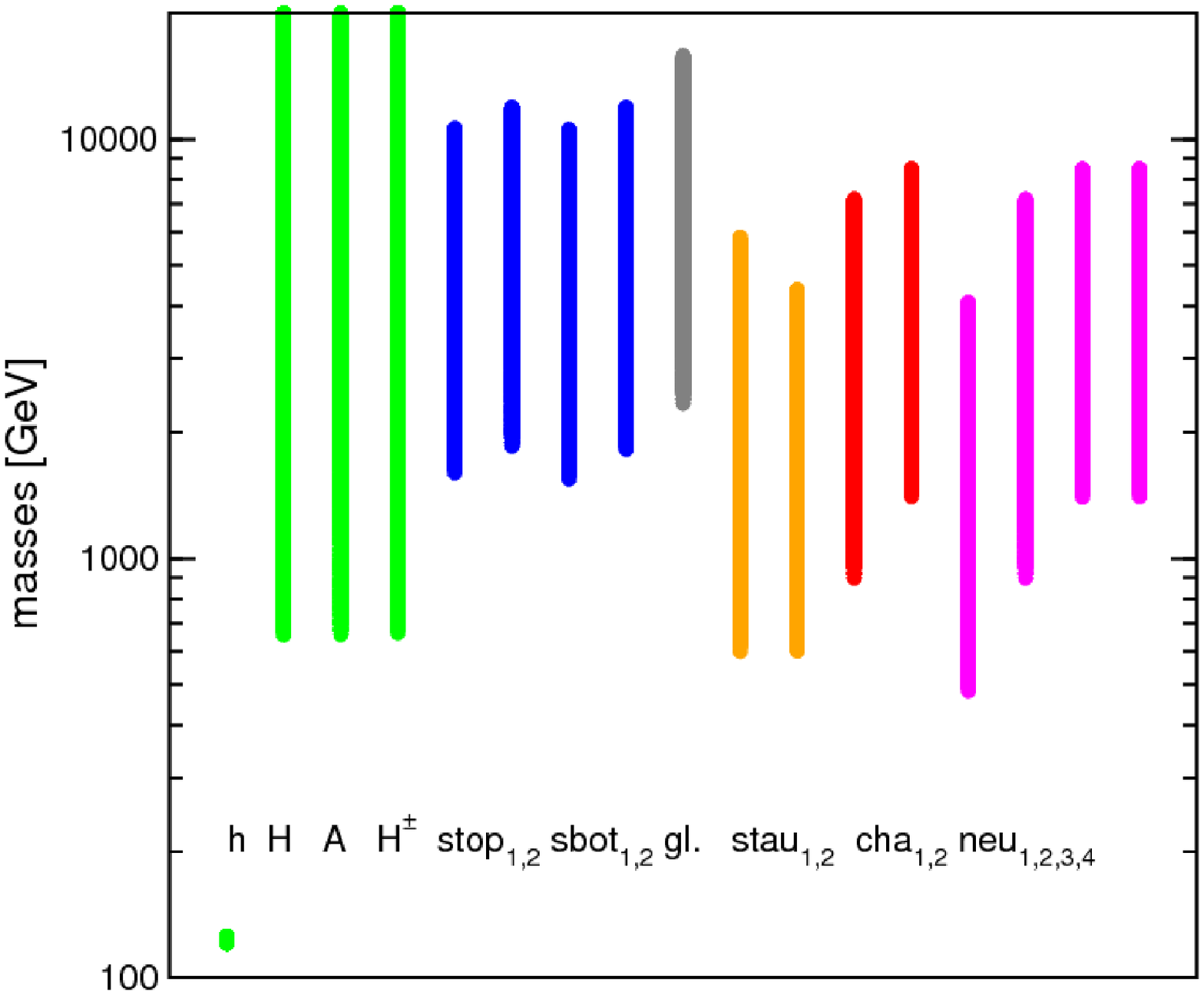}}
           \caption{The particle spectrum of model {\bf FUTB} with $\mu <0$,
             where 
             the points shown are in agreement with the quark mass
             constraints and the  
             $B$-physics observables. 
  The light (green) points on the left are the various Higgs boson masses. The
  dark (blue) points following are the two scalar top and bottom masses,
  followed by the lighter (gray) gluino mass. Next come the
  lighter (beige) scalar tau masses. The darker (red) points to
  the right are the two chargino masses followed by the lighter shaded (pink)
  points indicating the neutralino masses.}
\label{fig:masses}
\vspace{-0.5em}
\end{figure}%

The prediction of the lightest Higgs boson mass, evaluated with
FeynHiggs~\cite{Degrassi:2002fi,feynhiggs}, as a function of $M$ is 
shown in \reffi{fig:Higgs}. The light (green) points shown are in agreement
with all the constraints discussed listed above. 
The lightest Higgs mass ranges in 
\beq
M_h \sim 121-126~{\rm GeV} , 
\label{eq:Mhpred}
\eeq 
where the uncertainty comes from
variations of the soft scalar masses, and
from finite (i.e.~not logarithmically divergent) corrections in
changing renormalization scheme.  To this value one has to add $\pm 2$~GeV 
coming from unknown higher order corrections~\cite{Degrassi:2002fi}. 
We have also included a small variation,
due to threshold corrections at the GUT scale, of up to $5 \%$ of the
FUT boundary conditions.  
Thus, taking into account the $B$-physics constraints 
results naturally in a light Higgs boson that not only fulfills
the LEP bounds~\cite{Barate:2003sz,Schael:2006cr},
but naturally falls in the range favored by recent ATLAS and CMS
measurements~\cite{Dec13}. A more detailed discussion can be found in 
Refs.~\cite{Heinemeyer:2010xt,Heinemeyer:2007tz}.

The full particle 
spectrum of model {\bf FUTB} with $\mu <0$, again compliant with quark mass
constraints and the $B$-physics observables 
is shown in \reffi{fig:masses}. The masses of the particles increase with
increasing values of the unified gaugino mass $M$.

One can see that the lighter parts of the spectrum, especially of the colored
particles, are in the kinematic reach
of the LHC. A numerical example of such a light spectrum is shown in
Table~\ref{table:mass}. 
However, large parts are beyond this kinematic reach, and one possibility
offered by {\bf FUTB} with $\mu < 0$ would be to observe only a SM-like Higgs
boson around 125~GeV, but no additional SUSY particles.

\section*{Acknowledgments}
Dedicated to the memory of our
great friend and teacher in all respects Julius Wess.\\

This work is partially supported by the
NTUA's programme supporting basic research PEBE 2009 and 2010, and the
European Union's ITN programme ``UNILHC'' PITN-GA-2009-237920. Supported 
also by  mexican PAPIIT grant IN113712.
The work of S.H. is supported 
in part by CICYT (grant FPA 2010--22163-C02-01) and by the
Spanish MICINN's Consolider-Ingenio 2010 Program under grant MultiDark
CSD2009-00064. 





\end{document}